\documentclass[%
 reprint,
 amsmath,amssymb,
 aps,
 showpacs,
 showkeys
]{revtex4-1}

\usepackage{graphicx}
\usepackage{dcolumn}
\usepackage{bm}

\newcommand{\lsim}{\mathrel{\mathop{\kern 0pt \rlap
      {\raise.2ex\hbox{$<$}}}
    \lower.9ex\hbox{\kern-.190em $\sim$}}}

\newcommand{\gsim}{\mathrel{\mathop{\kern 0pt \rlap
      {\raise.2ex\hbox{$>$}}}
    \lower.9ex\hbox{\kern-.190em $\sim$}}}

\begin{document}

\preprint{arXiv:}

\title{On $^{146}$Nd, $^{144}$Sm and other unexplored 2$\beta$ decay isotopes}

\author{Francesco Nozzoli}
 \email{francesco.nozzoli@cern.ch}
\affiliation{%
  INFN-TIFPA, I-38123 Trento, Italy.
}%

\date{\today}

\begin{abstract}
  $^{146}$Nd is one of only four (over a total of 35 existing)
  2$\beta^-$ decay isotope candidates
  whose half-lives currently lack experimental limits.
  The $\alpha$ activity of the $^{146}$Sm
  daughter nuclide allows placement of limits on the 2$\beta$ decay
  using the $^{146}$Nd/$^{142}$Nd abundance ratio ($T^{\beta\beta}_{1/2} \gsim 3 \times 10^9$ yr)
  or direct search for $^{146}$Sm with accelerator mass spectrometers ($T^{\beta\beta}_{1/2} \gsim 4.5 \times 10^{19}$ yr).
  With a similar approach, a modest ($\sim$ Gyr) first limit on
  half-lives for the other unexplored $2\beta$ unstable isotopes
  and competitive limits (few $10^{13}$yr) for $^{98}$Mo and $^{122}$Sn are also given.
  Finally, it is shown how the
  limit $T^{0\nu\epsilon\beta^+}_{1/2} \gsim 10^{15}$ yr
  for the unexplored
  $^{144}$Sm $0\nu\epsilon \beta^+$ decay
  may be obtained from the data of a GSO crystal
  scintillator.
  
\end{abstract}


\maketitle


\section{Introduction}
Neutrinoless double beta decay ($0\nu2\beta$) process is currently
of great interest, since it is closely related to fundamental aspects
of elementary particle physics beyond the standard model \cite{PDG}.
In particular the investigation of $0\nu2\beta$ process
offers 
information complementary
to that given by neutrino oscillation experiments,
possibly revealing the nature of the neutrino (Majorana or Dirac particle)
and giving the absolute scale for the effective neutrino mass.

On the other hand, 2-neutrino double beta decay ($2\nu2\beta$)
is a second-order process,
which is not forbidden by lepton number conservation;
however, the investigation of this process in as many nuclei as possible
is very useful, since it gives information about the calculation
of nuclear matrix elements both for the $2\nu2\beta$ and for the
$0\nu2\beta$ processes \cite{TREV}.

Current experiments achieve impressive sensitivity to $0\nu2\beta^-$
and the existing limits on the half-life for this process are in
the $10^{24}-10^{26}$ yr range, making it possible to test for the effective
neutrino mass in the sub-eV range, depending on the considered isotope
(see e.g. \cite{CUORE,EXO,GERDA,KamLAND,NEMO,AURORA}).
Moreover, most of these experiments are able to identify the
$2\nu2\beta$ decay whose half-life lies in the $10^{19}-10^{21}$ yr range
for favorable isotopes (for a review see e.g. \cite{REV,BRB}).

Considering all of the
35 potential 2$\beta^-$ and 22 potential $\epsilon\beta^+$ decay candidates,
only eight nuclei still have no experimental
limits on half-life \cite{BRB,DBDtables95,DBDtables,bbrad};
this is generally due to a
low transition energy ($Q_{\beta\beta}$), a low natural abundance, or to
difficulties in obtaining a detector material containing the
2$\beta$ active nuclei.

\section{Limits from the abundance ratio in the Earth's crust}
A simple preliminary limit on the half-life of all unexplored 2$\beta$ decay
isotopes can be inferred from the daughter/parent abundance ratio in the Earth's crust \cite{CRC}.

In particular, assuming that daughter isotope, $D(t)$, is stable (or that
the known decay half-life is much longer than the Solar System lifetime),
and that the 2$\beta$ process is the dominant decay process for the parent
isotope, $P(t)$, the growth equation for the system is:
\begin{eqnarray}
  \frac{dD(t)}{dt}
  = \frac{P(t)}{\tau_{\beta\beta}}
  +
  \frac{dD^{ext}}{dt} \geq
  \frac{P(t)}{\tau_{\beta\beta}}
  \label{eq:eq1}
\end{eqnarray}
where $\frac{dD^{ext}}{dt}$ is the possible contribution coming
from sources that are different from the $2\beta$ decay of the parent isotope,
and this rate is positive since the daughter isotope is stable.
Equation \ref{eq:eq1} can be integrated from the time
of the Solar System formation ($t = 0$) to the present time \mbox{($t = T \simeq 4.5$ Gyr)} 
to obtain the half-life limit:
\begin{eqnarray}
D(T) > P(T) \left( e^{\frac{T}{\tau_{\beta\beta}}} -1 \right) 
\Rightarrow
T^{\beta\beta}_{1/2} >
\frac{\ln(2) \cdot T}{
  \ln \left(\frac {D(T)}
  {P(T)} +1\right)}
\nonumber
\end{eqnarray}
where it was assumed that additional
production of the parent isotope, $P^{ext}$, is negligible with respect to
the amount of parent isotope, $P(0)$, produced at Solar System formation time.

This geochemical limit for $2\beta$ decay is therefore based on the D/P ratio mesured in the Earth's crust \cite{CRC}.

The limit is very conservative,
since it is related to the requirement that
the daughter nuclei not be overproduced, with respect to the parent ones,
during the Solar System life because of the
$2\beta$ decay.

Table 1 summarizes the limits on $2\beta$ half-lives 
that can be obtained for some $2\beta$ unstable nuclei.
The limits obtained are very modest, ranging between a few $10^8$ yr and a few $10^9$ yr.
However, $2\beta$ processes for these nuclei are constrained here for the first time.

It is important to note that for the
$^{80}$Se -- $^{80}$Kr case,
losses of Kr gas from the Earth's crust to the atmosphere are possible \cite{man}, therefore
the ratio $\frac {^{80}\mathrm{Kr}}{^{80}\mathrm{Se}}$ may be underestimated in the 
Earth's crust
and the obtained limit might not be conservative.

On the other hand, since the $^{146}$Sm has a relatively short lifetime and ``immediately'' decays into $^{142}$Nd,
the limit of $^{146}$Nd $2\beta$ decay is related to the relative isotope abundance
of the same atomic species, which is
much better measured\textsuperscript{\footnote{The
  $\frac {^{146}\mathrm{Nd}}
  {^{142}\mathrm{Nd}}$ ratio is used as normalization in the Sm-Nd geochronology \cite{cron};
  some anomalies observed in the abundance ratios of Nd isotopes relative to $^{142}$Nd
  have been attributed to the possible contribution of $^{146}$Sm \cite{sm0,NATURE,geo}.
  Moreover,
  variations in the abundances of Nd isotopes in erupted lava are correlated with
  the volume of volcanic eruptions\cite{nat}.
}}
than the absolute abundance of different atomic species in the Earth's crust.

Finally, for $^{98}$Mo -- $^{98}$Ru and $^{122}$Sn -- $^{122}$Te $2\beta^-$ decay,
the current existing limits were calculated in \cite{DBDtables95} on the basis of the photographic emulsion measurements
of \cite{Fre52} with corrections on the decay energy and the natural abundance of the isotopes.
In the case of $^{98}$Mo -- $^{98}$Ru and $^{122}$Sn -- $^{122}$Te, the average $\frac {D(T)}{P(T)}$
ratios in Earth's crust are $6.5 \times 10^{-5}$ and $2.5 \times 10^{-4}$, respectively.
The limits $T^{^{98}\mathrm{Mo}}_{1/2} > 5 \times 10^{13}$ yr and
$T^{^{122}\mathrm{Sn}}_{1/2} > 1.3 \times 10^{13}$yr, obtained with 
the abundance ratio approach, are at the level of
the existing limits based on photographic emulsion analysis \cite{Fre52,DBDtables95}.

The approach described above, based on Earth's crustal abundance ratio, could be useful also to
set lower limits on half-lives for other unexplored rare/exotic nuclear decay processes \cite{sdense,pnc},
such as rare $\alpha$ decay \cite{caf2}, rare $\beta$ decay \cite{cd113,ca48} and cluster decay \cite{cdecay,lacl3}.

\section{The $^{146}$N\lowercase{d} $2\beta$ decay process}

Among all the 2$\beta^-$ decay candidates, $^{146}$Nd  (17.2 \% isotopic abundance) is
the nucleus having the lowest\textsuperscript{\footnote{Curiously, $^{150}$Nd
    (5.6\% isotopic abundance) provides the biggest $Q_{\beta\beta}$ value (3367.5 keV)
    among all $\beta^-$-stable 2$\beta^-$ decay candidates;
  measurements of $2\nu2\beta$ decay in $^{150}$Nd give:
  $T^{2\nu}_{1/2} = 8.2 \pm 0.9 \cdot 10^{18}$ yr
  \cite{BRB}.}}
$Q_{\beta\beta}$ value (70.2 keV) and it is still unexplored.
The theoretical prediction
for $2\nu2\beta^-$ decay of $^{146}$Nd gives the hopelessly-high
half-life $T^{2\nu}_{1/2} = 2.1 \cdot 10^{31}$ yr in the pseudo-SU(3) model \cite{PSU3};
this is due to the very low available phase space.
However, for the same reason (assuming
$\langle m_{\nu} \rangle  = 1$ eV), 
$0\nu2\beta^-$ decay of $^{146}$Nd
is expected to be a factor $\sim 10^3$
more probable than the standard $2\nu$ process \textsuperscript{\footnote{The $0\nu2\beta$ decay in $^{146}$Nd is expected to be dominant
  with respect to the $2\nu2\beta$ decay until
  $ \langle m_{\nu} \rangle  \gsim 0.03$ eV.}}.
In this latter case, in fact,
$T^{0\nu}_{1/2} = 1.18 \cdot 10^{28}$ yr is expected \cite{PSU3}.
Despite the very high expected half-lives, his feature
offers additional interest to the investigation of $^{146}$Nd double beta decay.

The isotope $^{146}$Nd is a potentially $\alpha$ radioactive nucleus
($Q_{\alpha} = 1.182$ MeV). However, since the expected half-life
for this decay channel is expected to be order $10^{34}$ yr \cite{YAG},
this contribution
can be neglected with respect to the $2\beta$ process.

The schema of $^{146}$Nd $2\beta$ decay to
$^{146}$Sm
is shown in Fig.1.
A peculiarity of the $^{146}$Nd 2$\beta^-$ decay is the $\alpha$ activity
of the daughter:
$^{146}$Sm
($T^{\alpha}_{1/2} = 68$ Myr,
\mbox{$Q_{\alpha} = 2529$ keV}) \cite{isot,tau}.
\begin{figure}[ht]
  \centering
  \includegraphics[width=0.5\textwidth]{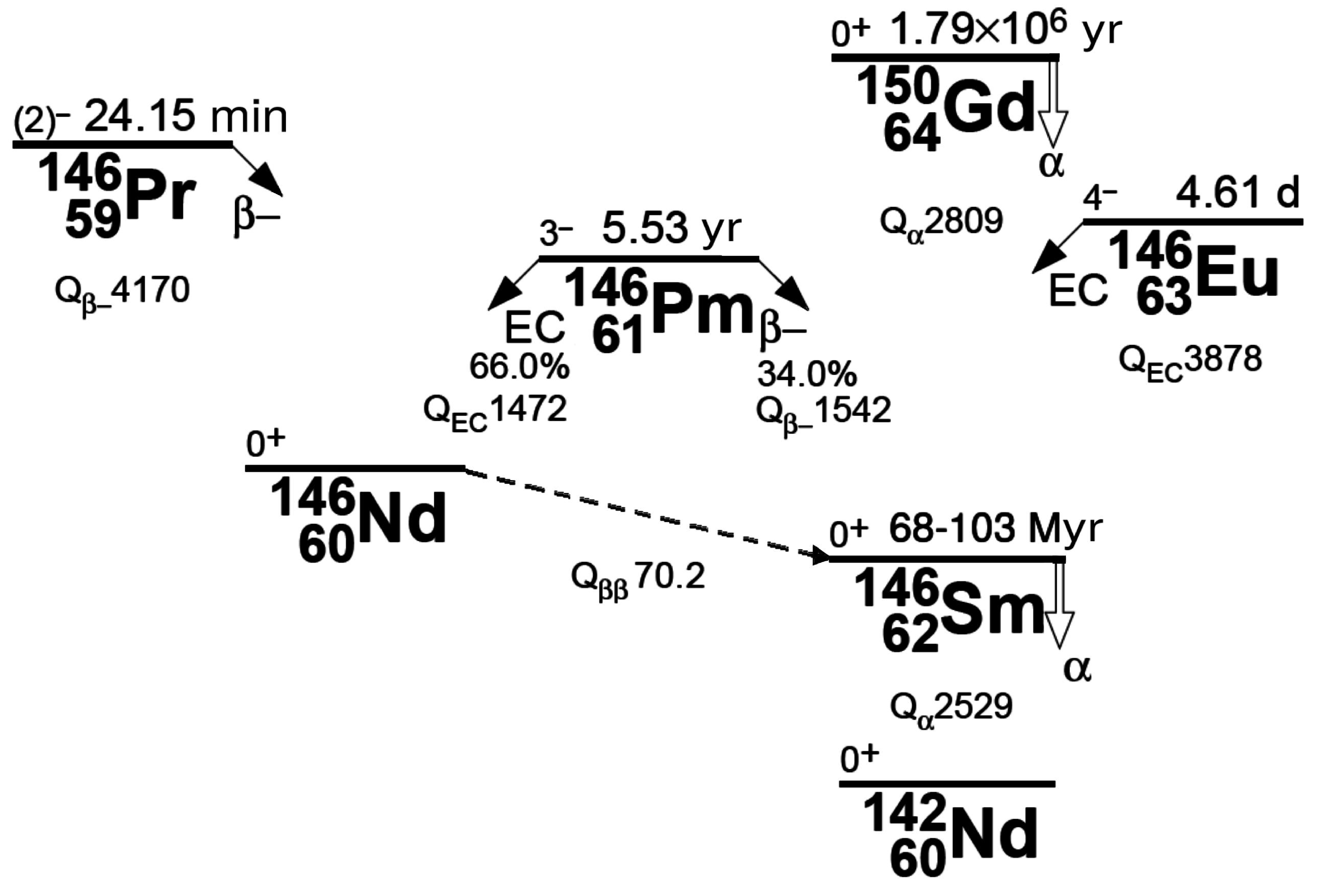}
  \caption{Simplified schema of $^{146}$Nd $\stackrel{2\beta^-}{\rightarrow}$ $^{146}$Sm
    $\stackrel{\alpha}{\rightarrow}$  $^{142}$Nd
    decay chain
    \cite{isot,tau}.
  }
\end{figure}

The half-life of $^{146}$Sm was recently re-determined in \cite{tau} to be 68$\pm$7 Myr, which is $\sim$30\% shorter than previous estimation (103$\pm$5 Myr) \cite{isot}.
Due to the relatively short half-life
of $^{146}$Sm,
the presence
of this isotope
in natural samples is expected to be very small.
Therefore, an alternative approach for the investigation of
$0\nu2\beta$ decay of $^{146}$Nd
is to search for the $^{146}$Sm $\alpha$ peak.
This approach, with respect to the direct investigation
of the $2\beta$ spectra, would be advantaged by
the reduction of the electromagnetic background
by possible pulse shape analysis and it would
avoid the difficulties posed by the small \mbox{$Q_{\beta\beta}$}.
On the other hand, despite its smallness,
the possible presence of relic $^{146}$Sm from
Solar System formation or from cosmogenic activation
would offer an unavoidable background contribution.

In fact, assuming equilibrium, the ratio of $2\beta$ produced $^{146}$Sm
over $^{146}$Nd is given by:
\begin{equation}
  \frac{
    ^{146}\mathrm{Sm}^{(\beta\beta)} }{^{146}\mathrm{Nd}}
  \simeq \frac{T^{\alpha}_{1/2}}{T^{0\nu}_{1/2}}
  \sim 10^{-20} \left( \frac{\langle m_\nu \rangle}{1 eV}\right)^2 \; .
  \label{eq:eq3}
\end{equation}
On the other hand, relic
$^{146}$Sm$^{\odot}$ from Solar System formation
is expected to be in the range:
\begin{equation}
  \frac{
    ^{146}\mathrm{Sm}^{\odot}(T)}{^{146}\mathrm{Nd}}
  =\frac{
    ^{146}\mathrm{Sm}^{\odot}(0)
  }
  {^{144}\mathrm{Sm}}
  e^
  {-\frac{T}{\tau_{\alpha}}}
  \frac{3.07\%}{17.2\%}
  \frac{\mathrm{Sm}}
       {\mathrm{Nd}}
       \sim 10^{-17} - 10^{-23}.
       \nonumber
\end{equation}
The large uncertainty in the latter evaluation is obtained by considering
the two more recent determinations of $^{146}$Sm lifetime:
\mbox{$\tau_{\alpha} = \frac{T^{\alpha}_{1/2}}{\ln2}\sim 100-150$ Myr}
 \cite{isot,tau}.
Moreover, the Solar System initial $\frac{^{146}\mathrm{Sm}^{\odot}(0)} {^{144}\mathrm{Sm}}$
ratio is expected to be of order $\sim 1\%$ \cite{sm0,tau}, and the ratio
$\frac{^{144}\mathrm{Sm}} {^{146}\mathrm{Nd}} = \frac{3.07\%}{17.2\%}
\frac{\mathrm{Sm}}{\mathrm{Nd}}\simeq 3\%$
\cite{CRC} has been considered.

Despite the possible high backgrounds provided by relic $^{146}$Sm$^{\odot}$
or by cosmogenic $^{146}$Sm production due to neutron spallation/capture,
a cautious lower limit on the $^{146}$Nd $2\beta$ decay half-life can be obtained.
It is important to note that the possible experimental detection of the
$^{146}$Sm ``background'' would be itself a very interesting physical result \cite{AMS},
giving additional information on $^{146}$Sm half-life, on
nuclear synthesis at the time of Solar System
formation \cite{sm0,EPSL,NATURE,tau}
and/or on the early crust-mantle differentiation process \cite{geo}.

As an example the limit of $^{146}$Sm/$^{147}$Sm$ < 10^{-11}$ was obtained in \cite{tau}
by analyzing the blank spectrum of $^{\mathrm{nat}}$Sm with the accelerator mass spectrometer (AMS).
Considering Eq. \ref{eq:eq3}, the limit $T^{^{146}\mathrm{Nd}}_{1/2}>4.5 \times 10^{19}$ yr can be inferred\textsuperscript{\footnote{
To infer the limit on $^{146}\mathrm{Nd}$ decay,
the assumption that Nd and Sm were not entirely separated
in natural samples must be considered. 
This is a reasonable assumption since  
Nd and Sm always occur together in nature because of their
similar chemical properties. 
As an example the Sm/Nd $\simeq 0.17$ measured in 
Earth's crust is very similar
to the Sm/Nd $\simeq 0.16$ measured in seawater \cite{CRC}.
The phenomenon of lanthanide contraction is mainly responsible
for this similarity \cite{AMS}.}}.

\section{Limits from S\lowercase{m}-doped detectors}

As an alternative to the search for $^{146}$Sm with AMS,
investigation of 2$\beta^-$ decay of $^{146}$Nd could be possible by a
direct search for the presence of the $^{146}$Sm $\alpha$ peak in detectors containing Sm.
Currently, there exists no appropriate Sm-based scintillator or bolometer.
A search for $^{146}$Sm could still be pursued in existing detectors
that have been doped with Sm or containing Sm as a contaminant.
As an example, a ZnWO$_4$ scintillating bolometer doped with enriched $^{148}$Sm
was used in \cite{isotta} to detect the $^{148}$Sm $\alpha$ decay
($T^{148}_{1/2} = 6.2^{+1.2}_{-1.2} \times 10^{15}$ yr and 
\mbox{$Q^{148}_{\alpha} = 1987.3 \pm 0.5$ keV}).
Despite the enrichment procedure, the faster $^{147}$Sm $\alpha$ decay
($T^{147}_{1/2} =1.06 \times 10^{11}$ yr) is clearly visible with a peak at 
\mbox{$Q^{147}_{\alpha} = 2310.5$ keV}.
On the other hand, just two events
are observed in the 2.5 MeV region where $^{146}$Sm is expected.
Isotopic composition measurements on enriched Sm$_2$O$_3$ powder are tabulated in \cite{isotta}.
After enrichment, the $^{147}$Sm was depleted by a factor 0.06 (from 14.99\% to 0.91\%)
and $^{144}$Sm was depleted by a factor 0.02 (from 3.07\% to 0.06\%). Considering the
enrichment of $^{148}$Sm by a factor 8.5 (from 11.24\% to 95.54\%), it is possible to
estimate that the depletion factor for the $^{146}$Sm should be in the range 0.03 - 0.04. 

Therefore, from the measurement reported in \cite{isotta}, a limit of $2 \times 10^{-8}$ on
$^{146}$Sm natural abundance can be inferred; this implies $T^{^{146}\mathrm{Nd}}_{1/2}>3.5 \times 10^{15}$ yr.
A similar limit ($\lsim 10^{-7}$) on the natural abundance of $^{146}$Sm can be obtained by
considering the spectrum of Li$_6$Eu(BO$_3$)$_3$ \cite{europium}.
Also in this case, the $^{147}$Sm $\alpha$ peak due to natural Sm
contamination is clearly visible but the $^{146}$Sm peak is absent. 
It is important to note that the limits obtained with this technique can be greatly improved if
a specific enrichment in $^{146}$Sm is pursued.
In particular, considering the ZnWO$_4$ used by \cite{isotta},
an improvement of a factor $\sim 250$ could be obtained by enriching the sample in
 isotope 146 instead of isotope 148. A furhter factor 100 improvement can be obtained
by a taking data for a period of a few years (the results of \cite{isotta} are based on an exposure of 364 h).
This would provide sensitivity to improve the AMS-based limit on $^{146}$Nd 2$\beta$ decay by a factor $\sim$10.

\section{A limit for $^{144}$S\lowercase{m} $0\nu\epsilon\beta^+$ decay}

$^{144}$Sm is one of only four $\epsilon\beta^+$
decay isotope candidates without experimental limits on half-life (see Table 1).
As above, this section discusses the possible sensitivity to 
$^{144}$Sm $0\nu\epsilon\beta^+$ decay offered by investigation
with detectors containing traces of Sm.
In particular, the concentration of $^{144}$Sm nuclei
in the detector can be inferred through the $\alpha$ activity of $^{147}$Sm.

Considering as an example the 635 g GSO scintillator of Ref. \cite{GSO}, the expected
contamination from Sm is at the level of $\sim$8 ppm \cite{GSO}; moreover,
since the GSO of Ref. \cite{GSO} is \mbox{$5.4$ cm $\times 4.7$ cm $\oslash$}, the detection
efficiency for an internal 511 keV $\gamma$ quanta is $\epsilon \gsim 0.5$
(for 511 keV in GSO, $\frac{\mu}{\rho} \sim 0.1$ cm$^2$/g).

Therefore, considering the limit $T^{^{160}\mathrm{Gd}}_{0\nu2\beta^-} > 1.3 \times 10^{21}$ yr
for $^{160}$Gd nuclei
given in Ref. \cite{GSO}
(which has a  $Q_{\beta\beta}^{^{160}\mathrm{Gd}} \simeq 1730$ keV very similar to $Q_{\beta\beta}^{^{144}\mathrm{Sm}} \simeq 1781$ keV)
a limit for $^{144}$Sm
nucleus $0\nu\epsilon\beta^+$ decay half-life
$T^{^{144}\mathrm{Sm}}_{0\nu\epsilon\beta^+} \gsim 10^{15}$ yr
can be deduced.

Finally, the sensitivity to $0\nu\epsilon\beta^+$ decay of $^{144}$Sm could be greatly improved
by requiring coincidences between near detectors, thanks to the
511 keV $\beta^+$-annihilation $\gamma$ quanta.
\begin{table*}[!hb]
  \caption{Half-life limits for unexplored (or poorly explored) 2$\beta$ processes;
    the average Earth's crust abundance of parent-daughter nuclei is considered \cite{CRC}.}
  \centering
  \begin{ruledtabular}
  \begin{tabular}{ l l l l l }
    \hline 
    Decay channel & Parent-Daughter & $\frac {N_{daughter}}{N_{parent}}$&
    $T^{\beta\beta}_{1/2}$ limit (yr)
    & Note \\
    $2\beta^-$ & $^{80}$Se -- $^{80}$Kr & $9 \times 10^{-5}$ & $3.5 \times 10^{13}$ &
    \footnote{Losses of Kr from the Earth's crust are possible \cite{man}.
      Limit is not conservative.} \\
    $2\beta^-$ & $^{86}$Kr -- $^{86}$Sr & $2 \times 10^{6}$ & $2 \times 10^{8}$ & \\
    $2\beta^-$ & $^{204}$Hg -- $^{204}$Pb & $33$ & $8.8 \times 10^{8}$ & \footnote{Daughter nucleus potentially $\alpha$ radioactive ($T^{daughter}_{\alpha} \geq T^{^{144}Nd}_{\alpha} = 2.3 \times 10^{15}$ yr \cite{DBDtables}).} \\
    $2\beta^- \rightarrow \alpha $ & $^{146}$Nd--$^{146}$Sm--$^{142}$Nd & $1.58$ & $3.3 \times 10^{9}$
    & \footnote{Parent nucleus potentially $\alpha$ radioactive ($T^{parent}_{\alpha} \geq T^{^{152}Gd}_{\alpha} = 1.1 \times 10^{14}$ yr \cite{DBDtables}).} \\
    $2\beta^- \rightarrow \alpha $ & $^{146}$Nd--$^{146}$Sm & -- & $4.5 \times 10^{19}$
    & $^{146}$Sm search in AMS \cite{tau}\\
    $2\beta^- \rightarrow \alpha $ & $^{146}$Nd--$^{146}$Sm & -- & $3.5 \times 10^{15}$
    & Sm doped ZnWO$_4$ \cite{isotta} \\

    $2\beta^-$ & $^{98}$Mo--$^{98}$Ru & $6.5 \times 10^{-5}$ & $5 \times 10^{13}$ &  $T^{\beta\beta}_{1/2}>10^{14}$ yr \cite{DBDtables95,Fre52}\\
    $2\beta^-$ & $^{122}$Sn--$^{122}$Te & $2.5 \times 10^{-4}$ & $1.3 \times 10^{13}$ & $T^{\beta\beta}_{1/2}>5.8 \times 10^{13}$ yr \cite{DBDtables95,Fre52}\\

    \hline
    $\epsilon\beta^+ + 2\epsilon$  & $^{144}$Sm -- $^{144}$Nd & $46$ & $8 \times 10^{8}$ & \footnotemark[2]
    \footnotemark[3] \\
    $0\nu\epsilon\beta^+$  & $^{144}$Sm -- $^{144}$Nd & -- & $10^{15}$ & Sm traces in GSO \cite{GSO}\\
    $2\epsilon$  & $^{152}$Gd -- $^{152}$Sm & $152$ & $6 \times 10^{8}$ & \footnotemark[3] \\
    $\epsilon\beta^+ + 2\epsilon$ & $^{162}$Er -- $^{162}$Dy & $273$ & $5.5 \times 10^{8}$ & \footnotemark[3] \\
    $2\epsilon$ & $^{164}$Er -- $^{164}$Dy & $26$ & $10^{9}$ & \footnotemark[3] \\
    $\epsilon\beta^+ + 2\epsilon$ & $^{168}$Yb -- $^{168}$Er & $226$ & $5.7 \times 10^{8}$ & \footnotemark[2]
    \footnotemark[3] \\
    $\epsilon\beta^+ + 2\epsilon$ & $^{174}$Hf -- $^{174}$Yb & $210$ & $5.8 \times 10^{8}$ & \footnotemark[2]
    \footnotemark[3] \\
  \end{tabular}
  \end{ruledtabular}
\end{table*}

\section{Conclusions}

Limits for half-lives of $2\beta$ processes in 10 isotopes (see table 1) are set here for the first time;
this completes the picture of $2\beta$ decay tables presented in Ref. \cite{DBDtables}.
The limits obtained from the absolute abundance ratio of elements in the Earth's crust
are generally very modest (at level of $\sim$ Gyr); however, for some elements,
such as $^{98}$Mo and $^{122}$Sn, this approach gives limits of a few $10^{13}$ yr that are at
the level of existing experimental ones.
Moreover, for  $^{146}$Nd and $^{144}$Sm isotopes, half-life limits of
$T^{\beta\beta}_{1/2} \gsim 4.5 \times 10^{19}$ yr and
$T^{0\nu\epsilon\beta^+}_{1/2} \gsim 10^{15}$ yr, respectively,
can be obtained from existing experimental data taken with accelerator mass spectrometers or
with Sm-containing detectors.

\begin{acknowledgments}
  I wish to acknowledge all the colleagues involved in useful discussions
  for this work and some other ones \cite{bilancia,disk}.
  \end{acknowledgments}

\bibliography{paper}

\end{document}